# Capacity Enlargement Of The PVD Steganography Method Using The GLM Technique

Mehdi Safarpour
Department Of Electrical Engineering
Zanjan University
Zanjan, Iran

Mostafa Charmi
Department Of Electrical Engineering
Zanjan University
Zanjan, Iran

*Abstract–In most steganographic methods, increasing in the capacity leads to decrease in the quality of the stego-image, so in this paper, we propose to combine two existing techniques, Pixel-value differencing (PVD) and Gray Level Modification (GLM), to come up with a hybrid steganography scheme which can hide more information without having to compromise much on the quality of the stego-image. Experimental results demonstrate that the proposed approach has larger capacity while it's results are imperceptible. In comparison with original PVD method criterion of the quality (PSNR) is declined by 2%(in decibel scale) averagely while the capacity is increased around 25%.*

I. INTRODUCTION

The modern digital world grows an interest in digital communications. This highly digitalized worlds has its particular pros and cons. One disadvantage of the this type of communications is security problems associated with it. Therefore researches have been done and techniques have developed to make the world a more secure place in terms of cyber security. These techniques mainly are classified into cryptography, steganography and watermarking methods [1].

Cryptographic methods often intend to hide data through deforming message to a meaningless message. The disadvantage of this is that an unusual message is noticeable and reveals existence of the secret communication. On the other hand steganographic techniques intent to transfer information through a media without anyone noticing even very existence of the secret communication.

Watermarking methods are developed to protect ownership rights of multimedia products, in fact these techniques are designed to protect the medium itself, so the amount of embedded data is not so important but instead robustness which means the immunity of the embedded data against attacks gains importance [2]. Unlike the watermarking methods, in the steganographic methods the amount of the embedded data has importance and the robustness comes second[3].

Steganographic methods work by embedding data in a medium such as an image or a video and, as mentioned above, they incline to not reveal even very existence of the secret communication. These methods does not change form of the secret message but disguise message through modifying the medium. Steganographic methods are mainly classified into spatial domain based and frequency domain based methods [1]. Among former we can mention the Pixel Value Differencing method and the Gray Level Modification method[5]. The PVD method proposed by Wu and Tsai [4] embeds a great deal of data and provides an imperceptible stego-image(the image that contains the secret data). The GLM technique proposed by Potdar et al[5], maps (and does not embed) secret information to the pixel values of the image. It exploits concept of the odd and even values to represent 0s and 1s of a secret bit stream. We will discuss both methods in subsequent sections, since in this work, we use the GLM method to enhance capacity of the PVD method and obtain a hybrid method.

Usually to evaluate embedding performance of a new scheme, two benchmarks are adopted. They are capacity of the embedded secret data and the quality of the stego-image. The capacity means how much data in bits or bytes can be embedded in an image using a supposed method. The quality of the stego-image reveals how much the images is distorted after embedding or how much the distortion of the image is perceptible to human eye and is estimated by the PSNR[1] measurement.

---

[1] Peak Signal to Noise Ratio

In this paper we will first discuss the PVD method and then we'll continue with the GLM technique and present our proposed method. Finally the empirical results and conclusion will be presented.

## II. BACKGROUND

### A. REVIEW OF THE PVD METHOD

Pixel value differencing method is a spatial domain steganographic method which can embed data through differences of adjacent pixels. In the original method published by Wu and Tsai, the grayscale image is divided into non-overlapping blocks containing two consecutive pixels, $(g_i$ and $g_{i+1})$. From each pair a difference, $d_i$, is calculated and classified into some ranges and based on the evaluated range some bits of binary stream of secret data are taken and converted to decimal format which is denoted by 'b'(this number will be embed in the difference of the pixels). Then new values of the pair are calculated and the new pixel values, $(g'_i, g'_{i+1})$, are obtained. After that the receiver of the message can retrieve the embedded part of the secret bit stream, 'b', by differencing this new assigned pixels. The whole algorithm is as follows:

Step 1 : The image is converted into a vector. The vector then is portioned to blocks of non-overlapping units ($g_i$ and $g_{i+1}$), then difference of each unit, $d$, is calculated by $d = g_{i+1} - g_i$.

Step 2 : The absolute difference $|d|$ is assigned into one of the classes of $R_k$. A practical setting of classes is $R_1=[0\ 7]$, $R_2=[8\ 15]$, $R_3=[16\ 31]$, $R_4=[32\ 63]$, $R_5=[64\ 127]$ and $R_6=[128\ 255]$. The width, the lower band and the higher band of $R_k$ are denoted by $w_k, l_k, u_k$ respectively.

Step 3: Length of embedded bit is determined by:

$$n = \lfloor log_2(w_k) \rfloor \qquad (1)$$

Then n bits of the secret data is selected and equivalent decimal value, 'b', is calculated.

Step 4: New differences are calculated by:

$$d' = \begin{cases} l_k + b, & d \geq 0 \\ -(l_k + b), & d < 0 \end{cases} \qquad (2)$$

and new pixel values are computed as follows:

$$(g'_i, g'_{i+1}) = \begin{cases} \left(g_i - \left\lceil \frac{d'-d}{2} \right\rceil, g_{i+1} + \left\lfloor \frac{d'-d}{2} \right\rfloor \right) & d \in odd \\ \left(g_i - \left\lfloor \frac{d'-d}{2} \right\rfloor, g_{i+1} + \left\lceil \frac{d'-d}{2} \right\rceil \right) & d \in even \end{cases} \qquad (3)$$

The following figure illustrates the embedding process for first pair pixels in an image. The numbers indicate gray level values of the pixels.

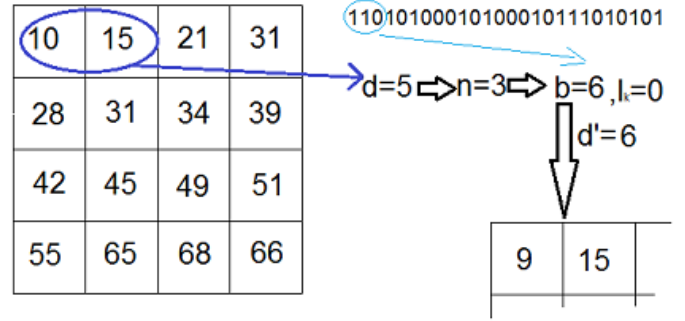

Figure 1. illustration of the PVD embedding process

To retrieve embedded data the absolute difference value of two consecutive pixels, $(g'_i, g'_{i+1})$, is calculated $(d'_i=|g'_{i+1}-g'_i|)$, then just same as hiding phase the related ranges($w_k, l_k, u_k$) of $d'$ are found. After that the embedded part of the secret data is calculated by subtracting $l_k$ from $d'$ as below:

$$b = |d'| - l_k \qquad (4)$$

By performing this process for each pair all of the secret message is retrieved[4].

### B. Steganography Using GLM Technique

In this subsection we discuss the Gray Level Modification algorithm for hiding secret data within the spatial domain of a gray scale image. The GLM technique is proposed by Potdar et al[5]. It employs concept of odd and even numbers for mapping data to the pixel values of an image. It assumes 1s are represented by odd values(pixels) and 0s are represented by even values(pixels). It transmits data through an image by mapping data to even and odd values of the gray level pixels. It modifies all pixels to make them even and odd to represent 0s and 1s of a secret bit stream. The GLM algorithm is described as follows [9]:

*1. Choose pixels in accord with an arbitrary mathematical function g (x, y).*
*2. Change the gray level values of the picked out pixels to make them even. These even gray levels will represent zero in a bit stream.*
*3. to represent 1, change the corresponded pixel by modifying its gray level value.*
*4. So we can represent both 1s and 0s using pixels.*

First step engages some pixels based on an arbitrary mathematical function which is common between the sender and the receiver. The second step make values of the selected pixels even. In third step based on the secret bit stream some of the selected pixels are made odd so we have a gray image with some even and odd pixels which are correspond to 0s and 1s of the secret bit stream.

To illustrate the procedure let's consider '101000100 0110110' bit stream as a sample secret data, the second and third steps of the algorithm which map the bit stream is shown by fig. 2 Note that we engage all pixels for mapping process and there is no selection function.

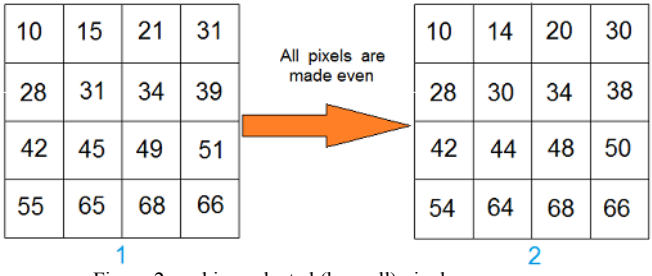

Figure 2. making selected (here all) pixels even

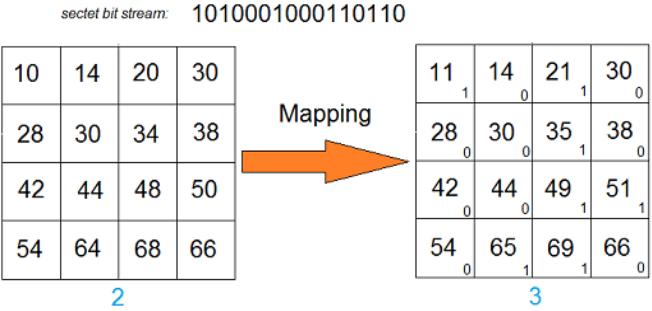

Figure 3. mapping ones and zeros by making some even values odd

To retrieve data the receiver picks up the selected pixels based on the common function and maps them to the corresponding binary data so as odd numbers are mapped to 1s and even numbers are mapped to 0s. This method does not affect the quality of the image because for mapping each bit just one bit of every pixel is modified. The Capacity of the GLM steganography methods depends on size of the image and utmost is equal to number of the pixels [8].

## III. PROPOSED METHOD

In this section we describe our proposed approach for achieving a hybrid method which increases capacity of the PVD method using the GLM technique.

Applying the PVD method leads to produce three types of pixel pairs, pairs whose pixels are even($g'_{i+1}$ and $g'_i$ even), pairs whose pixels are odd($g'_{i+1}$ and $g'_i \in$ odd), and pairs which just one of their pixels is odd and the other is even. The proposed scheme takes advantage from the first and second types. Our approach has two phases, in the first phase some secret data are embedded within the image by the PVD method which described in previous section, then in the second phase some other data are mapped by the GLM technique to the image. Since the second phase should not lead to lose the secret data which are embedded in the first phase we designed some measurements to avoid the interference. After finishing the PVD procedure we engage pairs which both of their pixels are odd or both of them are even (first and second types). We do not embed any further data within pairs which has just an odd or just an even pixel, but we mark them as abandoned pixels of the second phase using concept of the negative and non negative numbers. The whole procedure was indicated in fig.4.

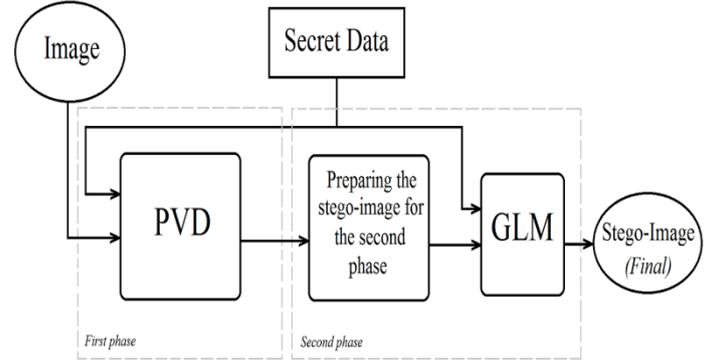

Figure 4. Whole procedure of embedding phase of the proposed method

Since the first phase is the PVD method and it discussed previously, we just describe second phase of the proposed method. The second phase is as follows:

*Step 1:* If ($g'_{i+1} - g'_i$) is negative, then change the spatial position of them with each other as indicated in (5). In this way all pairs will have a difference.

$$(g'_i, g'_{i+1}) = \begin{cases} (g'_i, g'_{i+1}) & d' \geq 0 \\ (g'_{i+1}, g'_i) & d' < 0 \end{cases} \quad (5)$$

We will use this trait to mark pairs which are not engaged in this phase of embedding process. Note that since the PVD embedded data are recovered using (4) the alteration of the sign of the differences, d', does not affect retrieving process of the first phase.

*Step 2:* Just in case both $g'_i$ and $g'_{i+1}$ are odd decrement them by one unit to make their values even (note that the $d'_i$ and the lower band $l_k$, which are needed to extract data in the first phase, does not change). If just one pixel of the pair ($g'_i$ or $g'_{i+1}$) is odd, change their position with each other in order to make their difference negative(in the previous step all differences were made, in this way we mark them as abandoned pairs to recognize them so we can know pairs which are not used to embedding data in the second phase).This step is represented in the following equation:

$$(g'_i, g'_{i+1}) = \begin{cases} (g'_i, g'_{i+1}) & g'_i, g'_{i+1} \in even \\ (g'_i - 1, g'_{i+1} - 1) & g'_i, g'_{i+1} \in odd \\ (g'_{i+1}, g'_i) & (g'_i + g'_{i+1}) \in odd \end{cases}$$

(6)

If both pixels are even, (6) does not change anything, if both are odd it makes them even (since $|g'_{i+1} - 1 - (g'_i - 1)| = |g'_{i+1} - g'_i| = |d'_i|$, the related difference does not change and the embedded data of the first phase still can be

retrieved). If just one of the pixels is odd, it changes spatial position the pixels so the related difference becomes negative.

*Step 3:* So we have some even value pixels. Just choose pairs with difference $(g'_{i+1} - g'_i \geq 0)$ and then based on the GLM technique and in accord with the secret bit stream make even values odd to represents 1s and do not change pixel when you want to represent 0s (the abandoned pairs are neglected in this phase).

First step makes all difference. To mark the abandoned pairs, we make their difference negative by changing their position with each other. Second step is somehow similar to the second step of the GLM method but it only modifies pairs whose both pixels are odd. Obviously the pairs whose both pixels are even do not need any alteration. Then according to the GLM technique some of these resulted even pixels( of none abandoned pairs) are made odd and are decremented by one to represent 1s and the others which should represent 0s do not change. Fig. 5-8 illustrate procedure of the proposed method.

| 8 | 15 | 21 | 31 |
|---|----|----|----|
| 28 | 31 | 34 | 39 |
| 45 | 37 | 21 | 50 |
| 29 | 20 | 27 | 35 |

PVD →

| 5 | 11 | 20 | 29 |
|---|----|----|----|
| 35 | 33 | 42 | 35 |
| 52 | 43 | 48 | 22 |
| 23 | 13 | 26 | 36 |

Figure 5. First phase, PVD embedding

Say the given block in the fig. 5 is a matrix of gray level values of an image(4×4). In this figure the PVD method was applied on the image to embed an arbitrary secret bits stream.

| 5 | 11 | 20 | 29 |
|---|----|----|----|
| 35 | 33 | 42 | 35 |
| 52 | 43 | 48 | 22 |
| 23 | 13 | 26 | 36 |

→

| 5 | 11 | 20 | 29 |
|---|----|----|----|
| 33 | 35 | 35 | 42 |
| 43 | 52 | 22 | 48 |
| 13 | 23 | 26 | 36 |

Figure 6. Applying first step

Equation (5) is applied on the resulted matrix to make all differences. Fig. 6. illustrates first step of the second phase.

| 5 | 11 | 20 | 29 |
|---|----|----|----|
| 33 | 35 | 35 | 42 |
| 43 | 52 | 22 | 48 |
| 13 | 23 | 26 | 36 |

→

| 4 | 10 | 29 | 20 |
|---|----|----|----|
| 32 | 34 | 42 | 35 |
| 52 | 43 | 22 | 48 |
| 12 | 22 | 26 | 36 |

Figure 7. Applying second step and the equation (6)

Fig. 7 illustrates the process of the second step. Consider for example the first pair,(5,11),since both of them are odd, we decrease both of them by one unit so they become even while their related difference does not change.(11-5=10-4=6) and embedded data of the first phase (PVD) still can be retrieved. Consider the second pair,(20,29),one of them is even and the other is odd so it should be marked as a abandoned pair so we changed their position in order to make the related difference negative and in this way we mark this pair as a abandoned pair.

Secret data:0111011101

| 4 | 10 | 29 | 20 |
|---|----|----|----|
| 32 | 34 | 42 | 35 |
| 52 | 43 | 22 | 48 |
| 12 | 22 | 26 | 36 |

→

| 4 | 11 | 29 | 20 |
|---|----|----|----|
| 33 | 35 | 42 | 35 |
| 52 | 43 | 22 | 49 |
| 13 | 27 | 26 | 37 |

Figure 7. Mapping further secret data by GLM technique to none abandoned pairs

As fig. 7 shows a secret bit stream is assumed to embed in the second phase which is shown on the top. The underlined pairs are in fact abandoned pairs which are not get involved in the second phase of embedding.

The retrieval process consists of two stages. Firstly embedded information of the second phase are extracted from none abandoned pairs by GLM retrieving process, then all odd values of the none abandoned pairs are made even again. Since the related differences are not changed, based on (4) rest of the data which were embedded in the first phase, are retrieved. Fig. 8 shows the retrieval process.

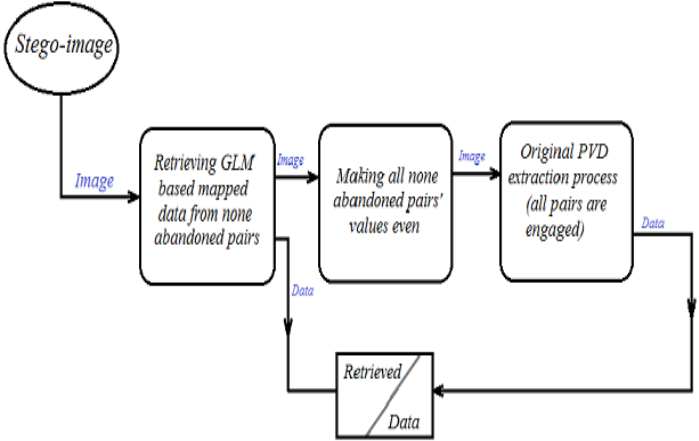

Figure 8. The retrieval process of the proposed algorithm

## IV. EXPERIMENTAL RESUALTS

In this section we present empirical results of the proposed method. To demonstrate performance of the method we applied the method on different images and compared the results with the related methods. In this paper we use four images namely "Lena" ,"Baboon" , "Peppers" and "Jet" which are mostly used images for evaluation of the steganographic schemes. Every image have been resized to 512×512 and converted to gray scale. Some examples of the cover image and its stego-image are shown in Fig. 9

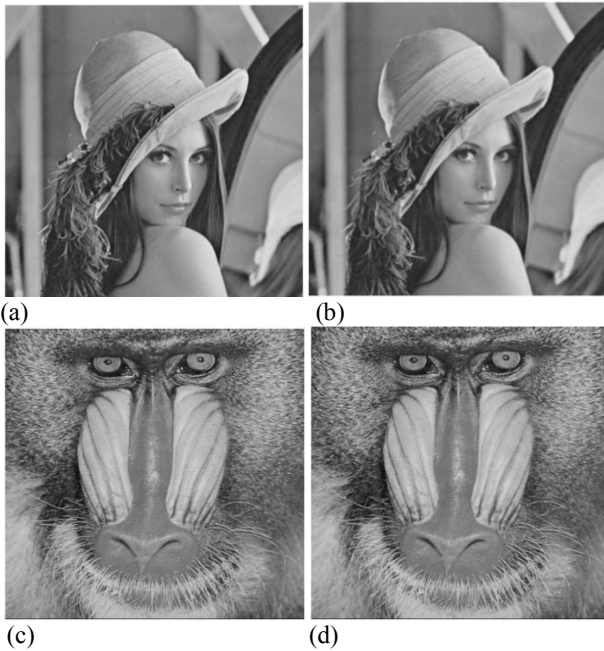

(a)  (b)  (c)  (d)
Figure 9  (a) Lena (b) Stego-image of Lena after embedding secret data with the proposed method (c) Baboon (d) Stego-image of Baboon after embedding secret data with the proposed method

To demonstrate invisibility of our proposed approach we have computed PSNR values.

### A. PSNR

To examine quality and imperceptibility of the stego-images, we used PSNR benchmark which reveals distortion of the embedded images. Peak Signal to Noise Ratio measurement is an evaluation of the quality[10]. To calculates the PSNR value,  the Mean Square Error should be calculated [11]. The MSE is computed as follows:

$$MSE = \frac{1}{mn}\sum_{i=0}^{m-1}\sum_{j=0}^{n-1}\|O(i,j) - S(i,j)\|^2$$

(7)

Where S and O are the stenographic image and original image respectively and m, n are the width and height of images. PSNR is calculated as follows:

$$PSNR = 10\log_{10}\left(\frac{MAX^2}{MSE}\right)$$

(8)

MAX is peak value in the signals and it is equal to 255 here, because it is the largest value in a gray scale image. Table I indicates PSNR values for different methods. The secret information bit stream is generated by random numbers. Fig. 10 indicates PSNR values for different sizes of "LENA" image.

Table I   COMPARISON OF THE QUALITY OF THE STEGO-IMAGE FOR THE PROPOSED METHOD AND OTHER SCHEMES

| Method \ Image | LENA | BABOON | PEPPERS | JET |
|---|---|---|---|---|
| PVD | 40.91 | 38.10 | 40.57 | 40.97 |
| GLM | 51.14 | 51.15 | 51.13 | 51.14 |
| **Hybrid PVD-GLM** | 40.05 | 37.21 | 39.12 | 40.12 |

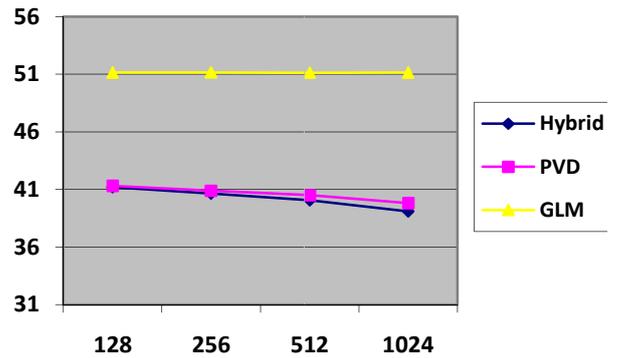

Figure 10 .PSNR for "LENA"

## B. Capacity

Here Capacity means the maximum bits that can be embedded in an image [1]. The proposed method shows high increment in the capacity. Table II shows calculated capacities of different methods. Increment of the capacity through our method is easily perceptive. Also fig. 11 shows a chart that indicates capacities for different sizes of "LENA" image.

Table II COMPARISON OF THE CAPACITY FOR THE PROPOSED METHOD AND OTHER SCHEMES

| Image  Method | LENA | BABOON | PEPPERS | JET |
|---|---|---|---|---|
| PVD | 406680 | 450328 | 405480 | 409944 |
| GLM | 262,144 | 262144 | 262144 | 262144 |
| **Hybrid PVD-GLM** | 526895 | 561649 | 617525 | 528898 |

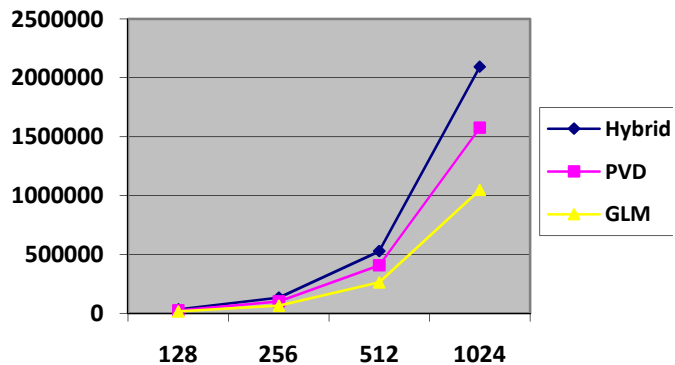

The empirical results show that utilizing GLM technique increases average capacity of the PVD method significantly (around 30%), while the average decrement of the PSNR value is inconsiderable and it is just around 2%. Since in this scheme after embedding by the PVD method every pixel are changed utmost by one unit, the PSNR value does not alter greatly.

## V. CONCLUSION

This paper introduces a new hybrid steganographic scheme which embeds more data and has a trivial effect on quality of the stego-image. It benefits from the GLM technique to improve the capacity of the PVD method. It has more embedding capacity than other methods while approximately does not reduce the quality of the stego-image. The experiments on the commonly used images demonstrate that proposed method improves the capacity of the PVD method around %25 while declines the PSNR value just around 2% (in decibel scale) so this method features the needed elements of the steganography and it is applicable for conveying secret information.